\documentclass[prd,aps,amsfonts,eqsecnum,nofootinbib,longbibliography,notitlepage,10pt]{revtex4-1}

\usepackage{graphicx}
\usepackage{xcolor}
\usepackage{rotating}
\usepackage{amsmath,amssymb,graphics,amsthm,isomath}
\usepackage{subfigure}
\usepackage{physics}
\usepackage{comment}

\usepackage[colorlinks=true, urlcolor=violet, linkcolor=blue, citecolor=red, hyperindex=true, linktocpage=true]{hyperref}
\usepackage[capitalise,compress]{cleveref}

\makeatletter
\renewcommand{\p@subsection}{}
\renewcommand{\p@subsubsection}{}
\makeatother

\usepackage{xcolor}
\usepackage{mathtools}

\usepackage{dsfont}

\begin{document}

\title{Non-Linear Dynamics and Critical Phenomena in the Holographic Landscape of Weyl Semimetals}

\author{Masataka Matsumoto}
\email{masataka@sjtu.edu.cn}
\affiliation{Wilczek Quantum Center, School of Physics and Astronomy, Shanghai Jiao Tong University, Shanghai 200240, China}

\author{Mirmani Mirjalali}
\email{mirmanimirjalali4@gmail.com}
\affiliation{Faculty of Physics, Shahrood University of Technology, P.O. Box 3619995161 Shahrood, Iran}

\author{Ali Vahedi}
\email{vahedi@khu.ac.ir}
\affiliation{Department of Astronomy and High Energy Physics,
Faculty of Physics, Kharazmi University, P. O. Box 15614, Tehran, Iran}

\begin{abstract}
This study presents a detailed analysis of critical phenomena in a holographic Weyl semi-metal (WSM) using the $D3/D7$ brane configuration. The research explores the non-linear response of the longitudinal current \( J \) when subjected to an external electric field \( E \) at both zero and finite temperatures. At zero temperature, the study identifies a potential quantum phase transition in the \( J \)-\( E \) relationship, driven by background parameters the particle mass, and axial gauge potential. This transition is characterized by a unique reconnection phenomenon resulting from the interplay between WSM-like and conventional nonlinear conducting behaviors, indicating a quantum phase transition.

Additionally, at non-zero temperature with dissipation, the system demonstrates first- and second-order phase transitions as the electric field and axial gauge potential are varied. The longitudinal conductivity is used as an order parameter to identify the current-driven phase transition. Numerical analysis reveals critical exponents in this non-equilibrium phase transition that show similarities to mean-field values observed in metallic systems.

\end{abstract}

\date{\today}
\maketitle
\tableofcontents
\section{Introduction}
Weyl semimetals (WSMs) have emerged as a fertile ground for studying various quantum phenomena due to their unique topological properties and gapless Weyl fermions \cite{Armitage2018}. WSMs are novel topological states of matter that exhibit linear energy-momentum dispersion relations and possess topologically protected band crossing points known as Weyl nodes \cite{Wan_2011,Burkov:2011ene,Hosur:2013kxa,Wehling_2014,Liu:2012hk,Lv_2015,Huang_2015,Xu_2015}. These materials exhibit unique transport properties, such as the chiral anomaly-induced negative magnetoresistance \cite{Son:2012bg,Burkov:2015hba,Li_2016,Liu:2012hk} and non-local transport \cite{Pellegrino_2015,Moll2016,Gorbar2018}. 

The critical behavior of these materials, particularly near phase transitions, is of significant interest in condensed matter physics. While equilibrium properties of WSMs have been extensively studied \cite{Madsen2021,Lahiri2019}, the non-equilibrium dynamics remain less understood, especially in the strongly coupled regime where traditional perturbative methods fail.

The study of quantum critical phenomena and non-equilibrium dynamics in condensed matter systems has been a subject of intense research in recent years. Of particular interest are materials exhibiting topological properties, such as WSMs, which possess exotic transport phenomena and quantum critical behaviors \cite{Wang219,Pixely2021,Li2018}. WSMs are characterized by the presence of Weyl nodes, which act as monopoles of Berry curvature in momentum space \cite{Berry_1984,Xiao_2010}. These nodes come in pairs of opposite chirality, and their separation in momentum space leads to interesting topological properties \cite{Armitage:2017cjs}.

The gauge/gravity duality, also known as the AdS/CFT correspondence, provides a powerful theoretical framework for probing the non-perturbative aspects of strongly coupled systems \cite{Maldacena1999, Witten1998, Gubser1998}. In this holographic approach, a gravitational theory in a higher-dimensional Anti-de Sitter (AdS) space is conjectured to be equivalent to a quantum field theory (QFT) on the boundary of this space and has been successfully applied to various condensed matter systems, including superconductors, superfluids, and strange metals \cite{Hartnoll:2009sz,Herzog:2009xv,Zaanen:2015oix}. In the context of holography, there have been several attempts to model WSMs using the gauge-gravity duality \cite{Landsteiner:2015pdh,Landsteiner:2015lsa,Copetti:2016ewq,Grignani:2016wyz,Ji:2021aan,Gursoy:2012ie}.

In this paper, we delve into the study of nonlinear response with respect to an external electric field in the holographic WSM using a probe brane approach. We build upon the top-down holographic model of WSM states constructed using $(3+1)$-dimensional $\mathcal{N}=4$ supersymmetric $SU(N_c)$ Yang-Mills theory, at large $N_c$ and strong coupling, coupled to a number $N_f \ll N_c$ of $\mathcal{N}=2$ hypermultiplets with mass $m$ \cite{BitaghsirFadafan:2020lkh}. In this work, we extend the analysis of holographic WSM using probe brane techniques to study the current-electric field ($J$-$E$) relation and the critical phenomena associated with non-equilibrium phase transitions driven by external currents. Our study is motivated by the need to understand the criticality that arises from the interplay between the topological nature of WSMs and the non-equilibrium conditions imposed by the nonlinear $J$-$E$ characteristics.

Our work contributes to the growing body of literature on critical phenomena and non-equilibrium dynamics in holographic WSMs. By studying the non-linear conductivity and current-driven phase transitions, we uncover novel aspects of the critical behavior and provide a deeper understanding of the underlying physics. Our findings have implications for both the fundamental theory of quantum criticality and the potential applications of WSMs in technological areas such as spintronics and quantum information processing.

The organization of this paper is as follows. In Section \ref{sec:rev}, we introduce the holographic model of a WSM, discuss the setup of the probe brane embedded in the bulk geometry, and review the conductivity of WSM from D3/D7. Section \ref{sec:non-linearity} is devoted to studying the characteristics of the non-linear response of current when exposed to an external field. In Sections \ref{sec:4} and \ref{sec:5}, we examine phase transition behaviors and critical phenomena in the critical point's threshold. Finally, following the discourse about our findings in Section \ref{sec:disc}, we finalize the paper with concluding remarks in Section \ref{sec:con}.

\section{Review on Weyl semimetals from D3/D7}\label{sec:rev}
We begin by studying a stack of $N_{f}$ D7-branes as a probe in the background of a stack of $N_c$ D3-brane, which at low energy provides the AdS black hole background:
\begin{equation}
	\dd s^{2} = \frac{L^{2}}{u^{2}}\left( -\frac{f(u)^{2}}{h(u)}\dd t^{2} + h(u) \dd \vec{x}^{2}\right) + \frac{L^{2}}{u^{2}} \dd u^{2} +L^{2}\dd\Omega_{5}^{2}, 
\end{equation}
where $f(u)=1-{u^4}/{u_{\rm H}^4}$ and $h(u)=1+{u^4}/{u^4_{\rm H}}$.
Here the black hole horizon is located at $u=u_{\rm H}$ and the AdS boundary is at $u=0$. The Hawking temperature is defined by $T=\sqrt{2} / (\pi u_{\rm H} L^{2})$.
The three-dimensional space coordinates are denoted by $\vec{x} = (x,y,z)$.
The metric of $S^{5}$ part in ten-dimensional spacetime is given by
\begin{equation}
	\dd\Omega_{5}^{2} = \dd\theta^{2} + \sin^{2}\theta \dd \psi^{2} + \cos^{2}\theta \dd \Omega_{3}^{2},
\end{equation}
where $\dd \Omega^{2}_{3}$ is the metric of $S^{3}$ part.
The dynamics of the system is obtained by the D7-brane action that is a sum of Abelian Dirac-Born-Infeld (DBI) and Wess-Zumino (WZ) terms,
\begin{equation}\label{action:0}
S_\mathrm{D7} = - N_f T_\mathrm{D7} \int d^8 \xi \sqrt{-\det (P[G] + 2\pi\alpha' F)}
+ \frac{(2\pi\alpha')^{2}}{2} N_f T_\mathrm{D7} \int P[C_4] \wedge F \wedge F,
\end{equation}
where $T_\mathrm{D7} = (2\pi)^{-7} g_s^{-1} \alpha'^{-4}$ is the D7-brane tension, $\xi^a$ with $a=1,2,\ldots,8$ are the worldvolume coordinates, $P[G]$ and $P[C_4]$ denote pullbacks of the bulk metric and 4-form to the worldvolume, respectively.
The U(1) gauge field strength is $2\pi \alpha' F = d A$ with the $U(1)$ worldvolume gauge field $A$.
Note that we have absorbed a factor of $2\pi\alpha'$ into the definition of the gauge field.
Here $C_4$ is an Ramond-Ramond 4-form potential
\begin{equation}
 C_4= \frac{L^4} {u^4} d t \wedge d x \wedge d y \wedge d z - L^4 \cos^4\theta d \psi \wedge \omega(S^3),
\end{equation}
where $\omega(S^3)$ denotes the volume form on a unit-radius $S^3$.
\begin{table}
	\centering
    \caption{Intersections of $D3$-branes and probe $D7$-branes.}
    \scalebox{1.2}{
{\tabcolsep = 1em	
 \begin{tabular}{|c|c|c|c|c|c|c|c|c|c|}
		\hline
		&$t$&$x$&$y$&$z$& $u$ &$S^3$&$\theta$& $\psi$ \\
		\hline 
		D3&$\times $&$\times$&$\times$&$\times$&  & & \hspace{1em} & \text{\hspace{1em}}\\
		\hline
		D7&$\times $ & $\times $&$\times$&$\times$ & $\times $&$\times$&  &\\
		\hline
	\end{tabular}
    \label{tab:1}
 }
 }
\end{table}
We assume that the D7-branes are extended along AdS$_5 \times S^3$ as shown in table \ref{tab:1}. We also, as is clear from table \ref{tab:1}, parameterize the D7-branes' worldvolume coordinates as $ (t,x,y,z,u)$ plus the $S^3$ coordinates.

For dealing with Weyl semimetals following \cite{BitaghsirFadafan:2020lkh}, we consider the ansatz for the scalar fields $\theta$ and $\psi$, where $\psi = bz + \phi(u)$ and $\theta=\theta(u)$.
With this assumption, the induced metric $P[G]$ is given by
\begin{equation}
	\dd s^{2}_\mathrm{D7} = \ \frac{L^{2}}{u^{2}}\left( -\frac{f(u)^{2}}{h(u)}\dd t^{2} + h(u) \dd \vec{x}^{2} \right)+L^2\left(\frac{1}{u^2}+\theta'(u)^{2} \right) \dd u^{2}+L^2\dd \Omega_{3}^{2},
 \label{eq:inducedmetric}
\end{equation}
and the pullback of $C_4$  becomes
\begin{equation}
P[C_4] = \frac{L^4}{u^4} \dd t \wedge \dd x \wedge \dd y \wedge \dd z - L^4 \cos^4\theta \left(b \, \dd z +  \frac{\partial \psi}{\partial u} \dd u \right) \wedge \omega(S^3).
\end{equation}
\\
To consider the conductivity in the Weyl semimetal, we assume that the gauge fields components are given by 
\begin{equation}
  A_x = - E t + a_{x}(u) , \quad 
  A_{y}= a_y (u),
\end{equation}
where $E$ is an external electric field.
With this assumption the D7-brane action \eqref{action:0} is written as 
	\begin{equation}
	\label{action:1}
	S_\mathrm{D7} = - \mathcal{N} V_{R^{1,3}} \int \dd u \left(\cos^3\theta(u)\,
	\sqrt{c_0(u)+c_1(u)  a_x'^2 +c_2(u) a_y'^2 +c_{3}(u) \phi'^{2}} -c_4(u) a_y'
	\right),
	\end{equation}
where
\begin{subequations}
	\begin{align}
		\label{action:coefficients:1}
		c_0&\equiv \left(b^2 g_{\phi \phi}+g_{xx}\right)\, \left(-E^2-g_{tt} g_{xx}\right) g_{uu}  g_{xx}\\
				c_1 &\equiv -g_{tt} g_{xx} \left(b^2
			g_{\phi\phi}+g_{xx}\right),\\
				c_2 &\equiv \left(b^2 g_{\phi\phi}+g_{xx}\right)
			\left(-E^2-g_{tt} g_{xx}\right) , \\
        c_3 & \equiv \left(-E^2-g_{tt} g_{xx}\right)g_{xx}^{2}g_{\phi\phi}\\
		c_4 &\equiv L^4 \cos^4\theta(u) \, b \, E.
	\end{align}
\end{subequations}
Also, ${\cal{N}}=N_{f}T_\mathrm{D7}(2\pi^{2})$ and $V_{R^{1,3}}$ is the volume of four dimensional spacetime. 
Hereafter, we will ignore the volume factor and consider (\ref{action:1}) as the action density.
As denoted in (\ref{eq:inducedmetric}), each component of the induced metric is explicitly written as
\begin{equation}
	g_{tt}=-\frac{L^2 f(u)^2}{u^2 h(u)}, \quad g_{xx}=\frac{L^2}{u^2}h(u), \quad
	g_{uu}=L^2 \left(\frac{1}{u^{2}}+\theta'(u)^{2} \right), \quad
	g_{\phi\phi}=L^2 \sin^2\theta.
\end{equation}
For simplicity, we set $L=1$ in the following.
Near the AdS boundary ($u=0$), each field can be expanded as\footnote{In the normalizable mode of $a_y$, we include the shift by $b E$ so that $j_y$ corresponds to the current density along the $y$-coordinate up to the constant factor.}
\begin{align}
\begin{aligned}
    \sin\theta(u) &= m u + c u^{3} + \cdots,\\
    a_{x}(u) &= a_{x}^{(0)} + \frac{j_x}{2} u^{2} + \cdots,\\
    a_{y}(u) &= a_{y}^{(0)}+\frac{j_y - bE}{2} u^{2} + \cdots,\\
    \phi(u) &= \phi^{(0)} + \frac{p_{\phi}}{2}u^{2} + \cdots,
    \label{eq:asympt}
\end{aligned}
\end{align}
where $m$ and $c$ are related to the quark mass $m_{q}$ and chiral condensate $\left<\bar{q}q \right>$ in the dual theory via the following relations:
\begin{equation}
    m_{q} = \frac{\sqrt{\lambda}}{2\pi} m , \quad \left<\bar{q}q \right>= 2 \frac{N_{c}\sqrt{\lambda}}{(2\pi)^{3}}  c,
\end{equation}
where $\lambda$ is the 't~Hooft coupling and now given by $\lambda = g_\mathrm{YM}^{2}N_{c}=4\pi g_s N_c$ with the relation $4\pi g_s N_c \alpha'^{2} = 1$.
In this paper, we refer to the geometric parameter $m$ as the quark mass.
We presume that the source terms for the gauge fields and pseudo-scalar field, denoted by $$(a_x^{(0)}, a_{y}^{(0)}, \phi^{(0)})$$ have become null.
Since the D7-brane action depends only on $(a_{x}',a_{y}',\phi')$ and not on $(a_{x},a_{y},\phi)$, we can define the corresponding constants of motion with respect to the $u$-coordinate. They are explicitly given by
\begin{eqnarray}
    \frac{\delta S_\mathrm{D7}}{\delta a_{x}'} &=& -{\cal{N}}\frac{c_{1}(u) \cos^{3}\theta \, a_{x}'}{\sqrt{c_0(u)+c_1(u)  a_x'^2 +c_2(u) a_y'^2 +c_{3}(u) \phi'^{2}}}, \label{eq:solax}\\
    \frac{\delta S_\mathrm{D7}}{\delta a_{y}'} &=& - {\cal{N}}\frac{c_{2}(u) \cos^{3}\theta \,a_{y}'}{\sqrt{c_0(u)+c_1(u)  a_x'^2 +c_2(u) a_y'^2 +c_{3}(u) \phi'^{2}}} +  {\cal{N}}c_{4}(u), \label{eq:solay}\\
    \frac{\delta S_\mathrm{D7}}{\delta \phi'} &=& - {\cal{N}}\frac{c_{3}(u) \cos^{3}\theta \,\phi'}{\sqrt{c_0(u)+c_1(u)  a_x'^2 +c_2(u) a_y'^2 +c_{3}(u) \phi'^{2}}}. \label{eq:solphi}
\end{eqnarray}
Considering the asymptotic forms of these conserved quantities near the AdS boundary with (\ref{eq:asympt}), one finds
\begin{equation}
    \lim_{u\to 0} \frac{\delta S_\mathrm{D7}}{\delta a_{x}'} = - {\cal{N}}j_x, \quad \lim_{u\to 0} \frac{\delta S_\mathrm{D7}}{\delta a_{y}'} = -{\cal{N}} j_y, \quad \lim_{u\to 0} \frac{\delta S_\mathrm{D7}}{\delta \phi'} = {\cal{N}} m^{2} p_\phi.
    \label{eq:currentdef}
\end{equation}
On the other hand, one can find the locus $u=u_{*} <u_{\rm H}$ at which $-E^{2}-g_{tt}g_{xx} = 0$ is satisfied, thus $c_0 = c_2 = c_3 =0$ there.
The locus of $u=u_{*}$ is referred as to the {\it effective horizon} because it is a causal boundary given by the effective metric on the worldvolume of the D7-brane \cite{Seiberg:1999vs,Kim:2011qh}.
Evaluating the above conserved quantities at the effective horizon, we obtain the expressions as a function of $u_{*}$:
\begin{equation}
    j_x = \sqrt{c_1 (u_{*})} \cos^{3} \theta(u_{*}), \quad j_y = -c_4(u_{*}), \quad p_{\phi} =0.
    \label{eq:currentform}
\end{equation}
Note that we find the trivial solution $\phi(u)=0$ because of $p_{\phi}=0$. 
According to the dictionary discussed in \cite{Karch:2007pd,BitaghsirFadafan:2020lkh}, these conserved quantities correspond to the dual operators associated with each respective field, namely\footnote{The sign of current densities is opposite to that defined in \cite{BitaghsirFadafan:2020lkh} because we used the opposite sign to define $j_x$ and $j_y$ in (\ref{eq:currentdef}).}
\begin{equation}
    \left< J_{x}\right> = (2\pi\alpha') {\cal{N}}j_{x}, \quad \left< J_{y}\right> = (2\pi\alpha') {\cal{N}}j_{y}, \quad \left< {\cal{O}}_{\phi}\right> =  {\cal{N}} p_{\phi},
\end{equation}
where $\left< J_{x}\right>$ and $\left< J_{y}\right>$ are the current density along $x$ and $y$ directions, and $\left< {\cal{O}}_{\phi}\right>$ is the pseudo-scalar condensate.
For simplicity, we refer to the quantities ($j_x, j_y,p_\phi$) as the current density and pseudo-scalar condensate by neglecting the constant factors.
In summary, the effective horizon is explicitly written as
\begin{equation}
    u_{*} =  \frac{\sqrt{2}}{\pi T} \sqrt{\frac{E}{(\pi T)^{2}} + \sqrt{1+\frac{E^{2}}{(\pi T)^{4}}}},
\end{equation}
and the current density along $x$ and $y$ directions are
\begin{eqnarray}
    j_x &=& \frac{f}{u_{*}^{2}}\sqrt{b^{2}\sin^{2}\theta + \frac{h}{u_{*}^{2}}}\cos^{3}\theta, \\
    j_y &=& - b E \cos^{3}\theta,
\end{eqnarray}
where each function is evaluated at $u=u_{*}$.
The Legendre transformation of \eqref{action:1} is

\begin{align}\label{action:final0}
	\tilde{S}_\mathrm{D7}  &\equiv  S_\mathrm{D7} - {\cal{N}}\int d u \left(  a_x' j_{x}  + a_y' j_{y} + \phi' p_{\phi}  \right)
	\nonumber \\
	&= - \mathcal{N} \int d u \,   \sqrt{
		c_0(u)\cos^6\theta -\frac{c_0 (u)}{c_3 (u)} p_{\phi}^{2} - \frac{c_0(u)}{c_1(u)}j_x^2 -\frac {c_0(u)}{c_2(u)}\left[j_y + c_4(u) \right]^2
	}, \nonumber\\
	\end{align}
where we used the solutions of ($a_{x}', a_{y}',\phi'$) derived from (\ref{eq:solax}) - (\ref{eq:solphi}).
Thus the action can be explicitly expressed as
\begin{equation}
        \tilde{S}_\mathrm{D7} = - \mathcal{N} \int \dd u \sqrt{g_{uu}g_{xx}}\sqrt{\left(-E^2-g_{tt}g_{xx}\right)\left(\cos^6\theta \left(b^2g_{\phi\phi}+g_{xx} \right)-\frac{j^2_x}{-g_{tt}g_{xx}}\right)-\left(j_y + b E \cos^{4}\theta \right)^2}, \label{eq:transformed-action}
\end{equation}
where we used $p_{\phi}=0$.
Here note that the reality condition of the action discussed in \cite{BitaghsirFadafan:2020lkh} presents the same expressions of conserved quantities as in (\ref{eq:currentform}).

In this paper, we are interested in the non-linear relation between the external electric field $E$ and longitudinal current $j_x$ with given background parameters $(T,m,b)$.
Since the system is invariant under the scale transformation, we will employ one of the background parameters to introduce scale-free quantities.
Concretely, we use $b$ in section \ref{sec:non-linearity}, \ref{sec:4} and $m$ in section \ref{sec:5} as a scaling parameter.

In the following calculations, we numerically solve the equation of motion for $\theta(u)$ obtained from the action (\ref{eq:transformed-action}) with each parameter given.
Here we do not show the explicit form of the ordinary differential equation because it is quite complicated and not very illuminating.
For the boundary conditions, we impose the regularity condition at the effective horizon and choose the value of $m$ given by the asymptotic behavior (\ref{eq:asympt}).
In the actual calculations, we fix $\theta(u_{*})=\theta_{0}$ and read off $m$ from (\ref{eq:asympt}) with the solutions since we numerically find solutions by shooting from the effective horizon.
By doing so with varying the parameters, we can obtain the relation between $E$ and $j_x$ for a desired value of $m$.

\section{Non-linear electrical conduction} \label{sec:non-linearity}
In this section, we study the non-linear behavior of the longitudinal current with respect to the external electric field both at zero and finite temperature.
Henceforth, we denote the longitudinal current density as $J\equiv j_{x}$ for simplicity because we focus only on the current density along the $x$ direction but not the $y$ direction.

\subsection{Zero temperature}

\begin{figure}[tbp]
    \centering
    \includegraphics[width=8cm]{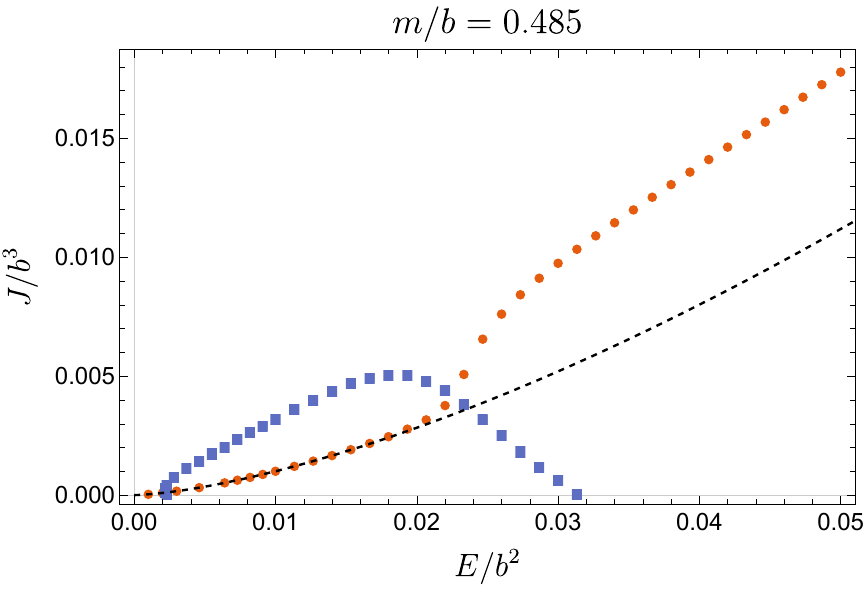}
    \includegraphics[width=8cm]{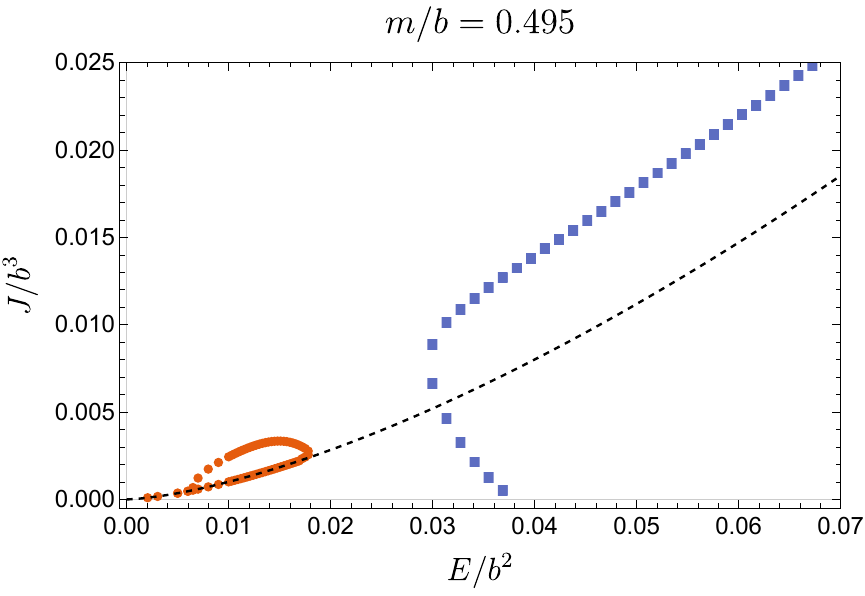}
    \caption{The relation between the longitudinal current $J/b^{3}$ and electric field $E/b^{2}$ for $m/b=0.485$ (left) and $m/b=0.495$ (right) at zero temperature.}
    \label{fig:JEplot}
\end{figure}

Figure \ref{fig:JEplot} shows the relations between the longitudinal current $J$ and the electric field $E$ for two different quark masses at zero temperature.
The black dashed curve denotes $J = E^{3/2}$ which is analytically obtained for small $E$ at zero temperature \cite{BitaghsirFadafan:2020lkh}.
For $m/b=0.485$ (left panel), one finds two branches: one is monotonically increasing with $E$ (circles), and the other exhibits non-linear behavior which includes the negative differential conductivity region (squares).
Remarkably, increasing the quark mass to $m/b=0.495$ (right panel), we find that these two branches are ``reconnected" and form new distinct branches that obviously show non-linear behaviors.
We consider that these two branches originate from the contribution of the WSM phase and dielectric breakdown. 
As the left branch in the right panel is well matched to the analytical form $J = E^{3/2}$ for small $E$, it corresponds to the metallic behavior in the WSM phase.
In the bulk picture, the solutions for small $E$ exist due to the presence of $b$, whereas a solution cannot be found for small $E$ with $b=0$ unless there is a finite charge density. 
In contrast, the right branch aligns with the conventional conducting phase achieved through dielectric breakdown, as initially observed in \cite{Nakamura:2010zd}. 
The gap for producing the current by the external electric field should be determined by the parameter $m/b$.
Therefore, this reconnection transition is certainly due to the competition between the WSM-like behavior and ordinary non-linear conducting behavior.
Our results could imply a kind of quantum phase transition between two different non-linear $J$-$E$ behaviors at a specific value of $m/b$, which is discussed later.

\subsection{Finite temperature}
\begin{figure}[tbp]
    \centering
    \includegraphics[width=12cm]{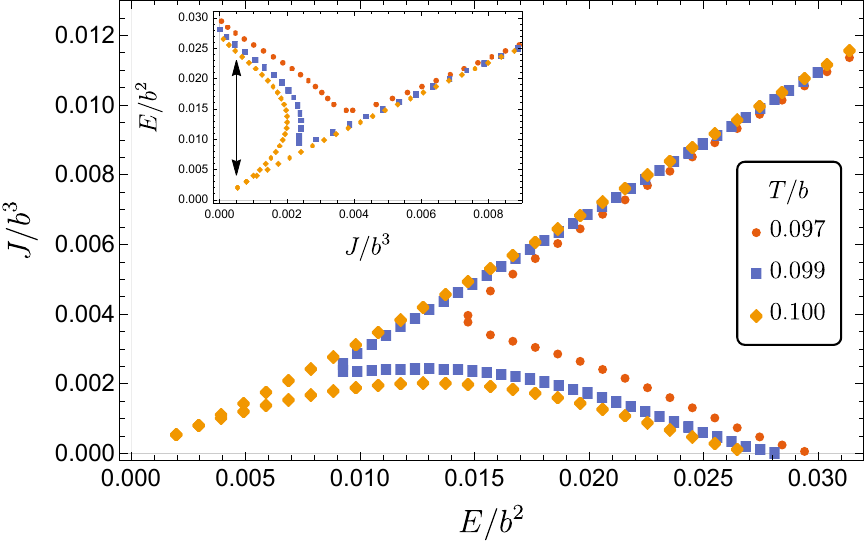}
    \caption{The relation between the longitudinal current $J/b^{3}$ and electric field $E/b^{2}$ for several temperatures with $m/b=0.5$ fixed. 
    The inset shows the same plot with the axes reversed.
    The arrow in the inset denotes the first-order phase transition point resulting from the thermodynamic analysis in section \ref{sec:4}.
    }
    \label{fig:JEplotT}
\end{figure}

Now we study the non-linear behavior of $J$ and $E$ at finite temperature.
In particular, we focus on the deformation of the $J$-$E$ characteristics in the ordinary conducting branch.
Figure \ref{fig:JEplotT} shows the $J$-$E$ behavior for several temperatures with the quark mass fixed $m/b=0.5$.
Note that in this parameter region, the WSM-like metallic behaviors in small $E$ are not found.

Here, let us consider that a fixed current drives the system, thus our control parameter is $J/b^{3}$ (see the inset of figure \ref{fig:JEplotT}).
For lower temperatures, such as $T/b=0.097$, only one value of $E/b^{2}$ corresponds to a fixed $J/b^{3}$.
If we continuously change $J/b^{3}$, the corresponding $E/b^{2}$ are continuously varied as if the crossover-like behavior.
However, as shown in the plot of $T/b=0.100$, the region where the multiple values of $E/b^{2}$ are possible at a given $J/b^{3}$ appears for a higher temperature.
Since the steady state that has only one of the $E/b^{2}$ values should be realized, the system is expected to undergo the first-order phase transition at a specific value of $J/b^{3}$. 
The critical temperature, defined by the transition point between the crossover and the first-order phase transition, is given by $T_{c}/b \approx 0.0987$.
Note that the $J$-$E$ curve reaches the origin for an even higher temperature, indicating the meson melting by temperature.
The deformation of the $J$-$E$ curve with respect to temperature has been observed in the ordinary conducting phase without $b$ \cite{Nakamura:2012ae,Ali-Akbari:2013hba,Matsumoto:2018ukk}, and our results qualitatively agree with them even for finite $b$.

\section{Current driven phase transitions and critical phenomena} \label{sec:4}

\subsection{Thermodynamic potential}
As mentioned in the previous section, the system undergoes the non-equilibrium phase transition driven by the current if we employ $J/b^{3}$ as a control parameter.
In this section, we investigate the critical phenomena in the vicinity of the critical point.
To do so, we need to determine the first-order phase transition point at which the $E/b^{2}$ value jumps when we continuously change $J/b^{3}$.

Following the discussion on the thermodynamic potential in the current-driven system \cite{Nakamura:2012ae,Ali-Akbari:2013hba,Matsumoto:2018ukk}, we define the thermodynamic potential with the small cutoff at the boundary ($u=\epsilon$) as
\begin{equation}
    F(T,J) \equiv \lim_{\epsilon \to 0} \left[ \int^{u_{*}}_{\epsilon} \dd u \bar{\cal{H}}_\mathrm{D7}   - S_{\rm ct}(\epsilon) \right],
\end{equation}
where $\bar{\cal{H}}_\mathrm{D7}$ is the Hamiltonian density, defined as 
\begin{equation}
    \bar{\cal{H}}_\mathrm{D7} = \dot{A_{x}}\frac{\partial \bar{\cal{L}}_\mathrm{D7}}{\partial \dot{A_{x}}} - \bar{\cal{L}}_\mathrm{D7},
\end{equation}
and assume that this quantity determines the thermodynamic stability of the system with a constant electric current given\footnote{In the probe brane model another definition of the free energy has also been proposed in \cite{Kundu:2013eba,Kundu:2019ull}.}.
Note that $\bar{\cal{L}}_\mathrm{D7}$ is not the same as $\tilde{\cal{L}}_\mathrm{D7}$ defined as $\tilde{S}_\mathrm{D7}=\int \dd u \tilde{\mathcal{L}}_\mathrm{D7}$ but the Lagrangian density we perform the Legendre transformation as
\begin{equation}
    \bar{\cal{L}}_\mathrm{D7} \equiv {\cal{L}}_\mathrm{D7} - \frac{\partial {\cal{L}}_\mathrm{D7} }{\partial A_{x}'} A_{x}' - \frac{\partial {\cal{L}}_\mathrm{D7} }{\partial A_{y}'} A_{y}',
\end{equation}
where $S_\mathrm{D7}=\int \dd u \mathcal{L}_\mathrm{D7}$.
Also $S_{\rm ct}$ denotes the counterterms to regularize the divergence at the boundary $u=0$.
The counterterms are divided into 
\begin{equation}
    S_{\rm ct} = \left(\sum_{i=1}^{5} S_{i} \right) - L_{F},  
\end{equation}
where each term is defined by \cite{Karch:2005ms,Karch:2007pd,BitaghsirFadafan:2020lkh}
\begin{eqnarray}
    S_{1} &=& \frac{1}{4} {\cal{N}} \sqrt{- \gamma} , \\
    S_{2} &=& -\frac{1}{2} {\cal{N}}\sqrt{- \gamma} \abs{\Theta}^{2}, \\
    S_{3} &=& \frac{5}{12} {\cal{N}}\sqrt{- \gamma} \abs{\Theta}^{4}, \\
    S_{4} &=& \frac{1}{2} {\cal{N}}\sqrt{- \gamma} \Theta^{*} \Box_{\gamma}\Theta \log \abs{\Theta}, \\
    S_{5} &=& \frac{1}{4} {\cal{N}}\sqrt{- \gamma} \Theta^{*} \Box_{\gamma}\Theta, \\
    L_{F} &=& -\frac{1}{4} {\cal{N}} \sqrt{- \gamma} F^{\mu\nu}F_{\mu\nu}\log \kappa \epsilon.
\end{eqnarray}
Here, $\Theta = \theta e^{i \phi}$ and $\gamma_{\mu\nu} = \epsilon^{-2} \eta_{\mu\nu}$ is the induced metric on the $u=\epsilon$ slice.
Note that the operator $\Box_{\gamma}$ is given by
\begin{equation}
    \Box_{\gamma} \Theta = \frac{1}{\sqrt{-\gamma}} \partial_{\mu} \left( \sqrt{-\gamma} \gamma^{\mu\nu} \partial_{\nu} \Theta\right),
\end{equation}
with $\gamma = \det \gamma_{\mu\nu}$.
In the last counterterm, $\kappa$ is a dimensionful constant to make the argument of the logarithm dimensionless.
In our study, we simply choose $\kappa = 1$ as employed in \cite{Nakamura:2012ae,Matsumoto:2018ukk,Jensen:2010vd}.
Evaluating these counterterms near $u=\epsilon$, we find that they are explicitly written as
\begin{equation}
    S_{\rm ct} = {\cal{N}} \left[ \frac{1}{4\epsilon^{4}} - \frac{m^{2}}{2\epsilon^{2}} - \left(b^{2}m^{2} + \frac{E^2}{2} \right)\log \epsilon - m \theta_{2} + \frac{5}{12}m^{4} - \frac{b^{2}m^{2}}{4} - \frac{b^{2}m^{2}}{2}\log m \right].
\end{equation}

Computing the thermodynamic potential as a function of $J/b^{3}$, we find that the solutions with smaller $E/b^{2}$ are thermodynamically stable compared to other solutions.
Thus, we consider that the first-order phase transition occurs at the left edge of the multi-valued region with respect to the current density, as denoted by the arrow in the inset of figure \ref{fig:JEplotT}.

\subsection{Critical phenomena}
\begin{figure}[tbp]
    \centering
    \includegraphics[width=10cm]{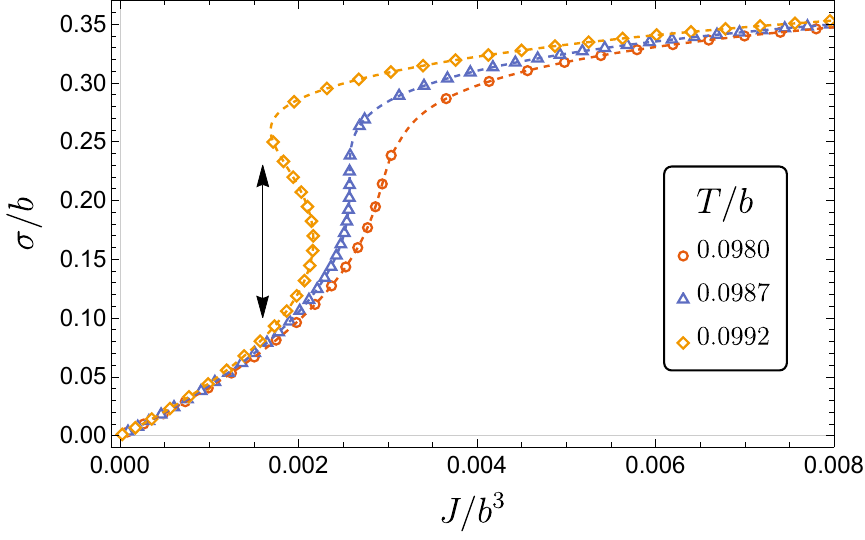}
    \caption{The longitudinal conductivity $\sigma/b$ as a function of the current density $J/b^{3}$ for several temperatures with $m/b=0.5$ fixed.
    The arrow denotes the first order phase transition point determined by the thermodynamic potential.}
    \label{fig:sigmaplot}
\end{figure}

Following the previous studies \cite{Nakamura:2012ae,Matsumoto:2018ukk}, we employ longitudinal conductivity $\sigma = J/E$ to define the order parameter for our current-driven phase transitions.
We also regard the current density $J$ as an external field for inducing the order parameter.
We show the conductivity as a function of the current density for several temperatures in figure \ref{fig:sigmaplot}.
As expected from the behaviors in figure \ref{fig:JEplotT}, by changing the current density the longitudinal conductivity also obviously undergoes second-order phase transition at $T_{c}/b\simeq 0.0987$, first-order phase transitions for $T>T_{c}$, and crossover for $T<T_{c}$.

\begin{figure}[tbp]
    \centering
    \includegraphics[width=8cm]{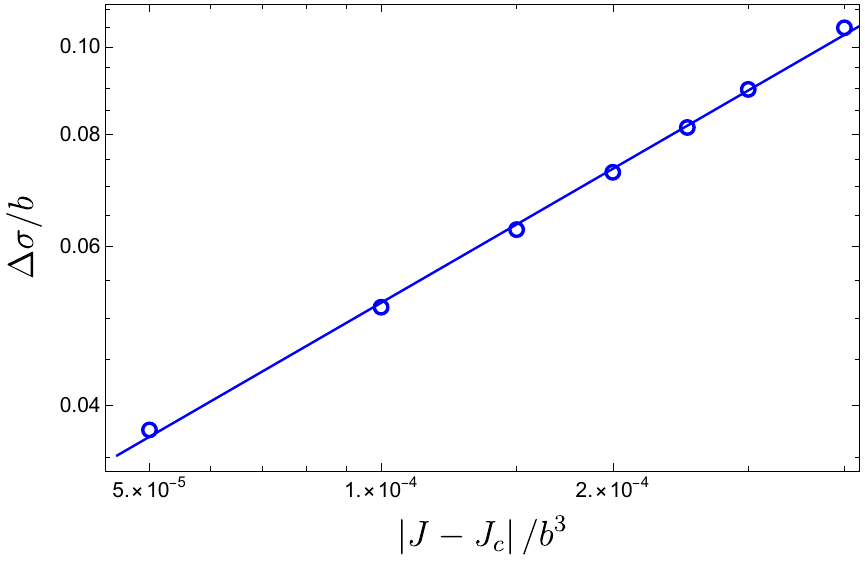}
    \includegraphics[width=8cm]{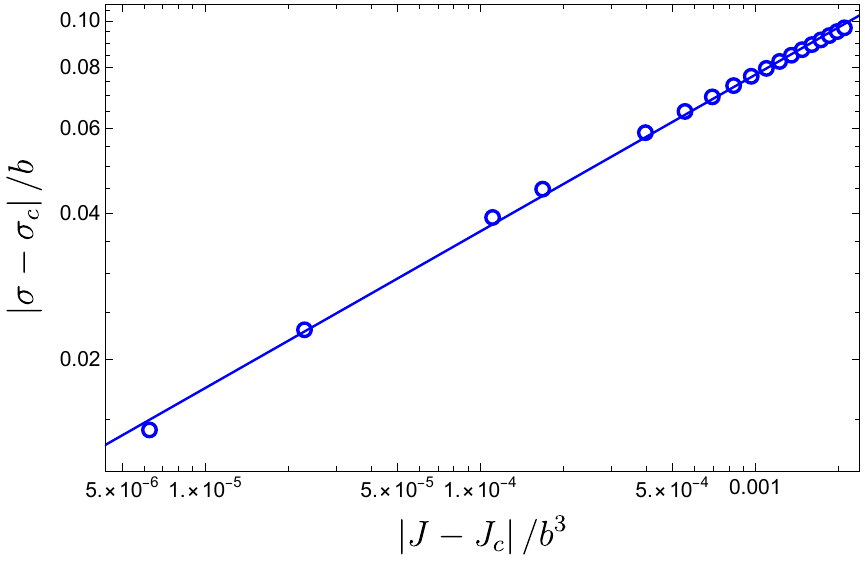}
    \caption{The critical behaviors associated with the critical exponent $\beta$ (left) and $\delta$ (right) in log scale. The solid lines show the numerical fitting results}
    \label{fig:exponent}
\end{figure}

Now we define the critical exponents as
\begin{eqnarray}
    \Delta \sigma &\propto& \left|  T- T_{c}\right|^{\beta}, \quad \left( T> T_{c}\right)  \label{eq:betadef}\\
    \left|\sigma -\sigma_{c} \right|  &\propto& \left| J-J_{c}\right|^{1/\delta}, \quad \left( T=T_{c}\right)
\end{eqnarray}
where $\sigma_{c}$ and $J_{c}$ are the critical values of conductivity and current density, respectively.
The exponent $\beta$ is related to the behavior of the order parameter, corresponding to the discontinuous jump of the conductivity: $\Delta\sigma \equiv \sigma_{+} - \sigma_{-}$, where $\sigma_{\pm}$ denotes the larger(smaller) value of the conductivity at the first-order phase transition point.
The exponent $\delta$ characterizes the relation between the order parameter and the external field at the critical point.
We show the numerical results of these critical exponents with $m/b=0.5$ fixed in figure \ref{fig:exponent} and we obtain
\begin{equation}
    \beta \approx 0.4941, \quad \delta \approx 3.087.
\end{equation}
These values approximately agree with those in the Landau theory of equilibrium phase transitions, that is
\begin{equation}
    \beta = \frac{1}{2}, \quad \delta =3, \label{eq:MFvalue}
\end{equation}
which corresponds to the critical exponents in the mean-field approximation.
This agreement was originally observed in the metallic setup \cite{Nakamura:2012ae,Matsumoto:2018ukk}.
Consequently, we find that the presence of axial gauge potential $b$ does not influence the critical phenomena of the current-driven phase transition observed in the D3/D7 model.

\section{\texorpdfstring{$b$}{TEXT} dependence of conduction and critical phenomena} \label{sec:5}
So far we have studied the non-linear $J$-$E$ characteristics by scaling the quantities with $b$.
In studying the critical phenomena of current-driven phase transitions, we chose temperature as a parameter that measures the ``distance" to the critical point, as in (\ref{eq:betadef}), in analogy with equilibrium phase transitions.
In this section, we use $m$ as a scaling parameter instead of $b$ and investigate the $J$-$E$ characteristics and critical phenomena.
Since the only difference is the choice of parameter for scaling, the calculation method is the same as in the previous section.
Therefore, we briefly show only the results in the following.

\begin{figure}[tbp]
    \centering
    \includegraphics[width=12cm]{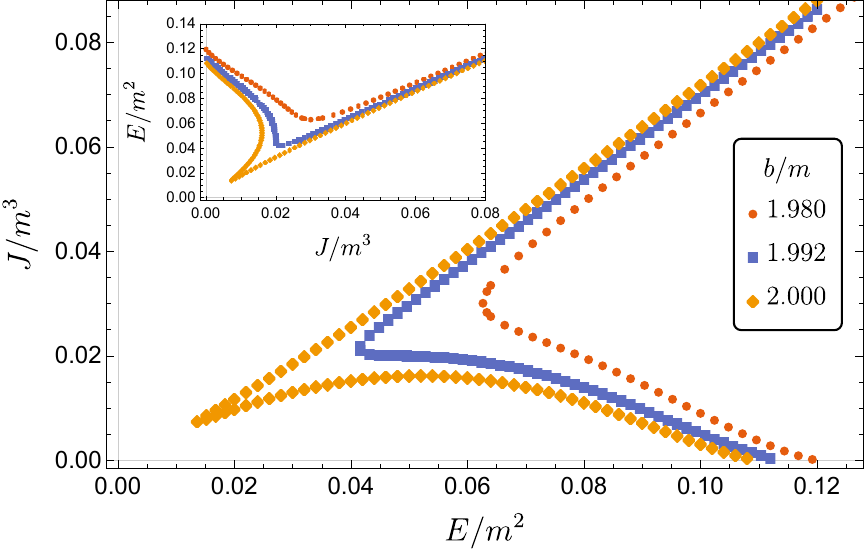}
    \caption{The relation between the longitudinal current $J/m^{3}$ and electric field $E/m^{2}$ for several values of $b/m$ with $T/m=0.2$ fixed. 
    The inset shows the same plot with the axes reversed.}
    \label{fig:JEplotb}
\end{figure}

Figure \ref{fig:JEplotb} shows the $J$-$E$ characteristics for several values of $b/m$ with $T/m=0.2$ fixed.
For larger $b/m$, as observed in figure \ref{fig:JEplotT} for larger $T/b$, $E/m^{2}$ becomes multi-valued with respect to a given $J/m^{3}$.
In the same manner as in section \ref{sec:non-linearity}, we determine the critical point of the current-driven phase transition, defined by the middle point between the crossover behavior at smaller $b/m$ and first-order phase transition at larger $b/m$, and obtain $b_{c}/m \approx 1.993$.

\begin{figure}[tbp]
    \centering
    \includegraphics[width=8cm]{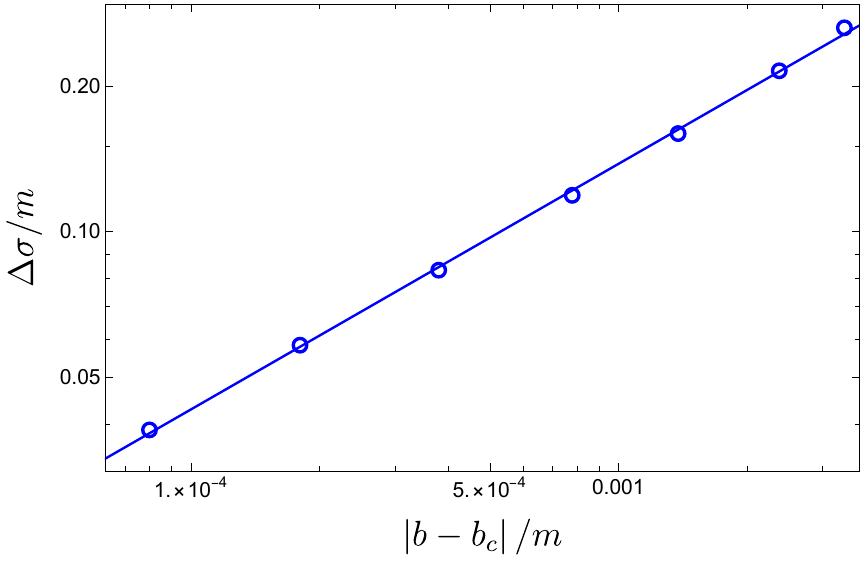}
    \includegraphics[width=8cm]{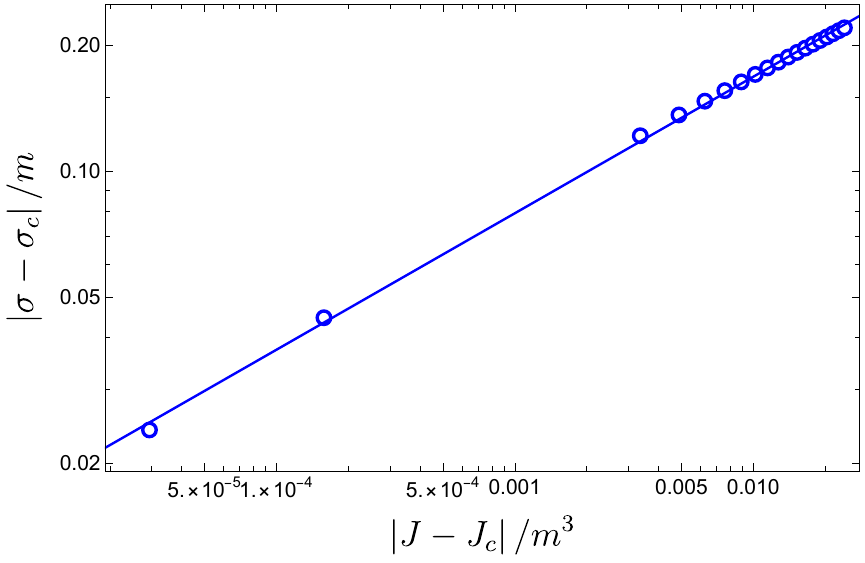}
    \caption{The critical behaviors associated with the critical exponent $\beta$ (left) and $\delta$ (right) in log scale. The solid lines show the numerical fitting results.}
    \label{fig:exponent_b}
\end{figure}

Now we define the critical exponents as
\begin{eqnarray}
    \Delta\sigma &\propto& \left| b-b_c \right|^{\tilde{\beta}}, \quad \left( b> b_{c}\right) \\
    \left|\sigma - \sigma_{c} \right| &\propto& \left|J-J_{c} \right|^{1/\tilde{\delta}}. \quad \left( b= b_{c}\right)
\end{eqnarray}
We show the critical behaviors associated with those critical exponents in figure \ref{fig:exponent_b}.
Here note that $b$ is the parameter that measures the distance to the critical point instead of $T$ in (\ref{eq:betadef}). 
As a result, we numerically obtain 
\begin{equation}
    \tilde{\beta} \approx 0.5076, \quad \tilde{\delta} \approx 0.3268,
\end{equation}
which again agree with the mean-field values (\ref{eq:MFvalue}).
Consequently, we find that the non-linear behaviors of $J$ and $E$ as well as the critical phenomena of the current-driven phase transitions are qualitatively similar irrespective of our choice of parameters.
Our results indicate that the axial gauge potential $b$ may act as thermodynamic variables, like the temperature, in the critical phenomena.

\begin{figure}[tbp]
    \centering
    \includegraphics[width=12cm]{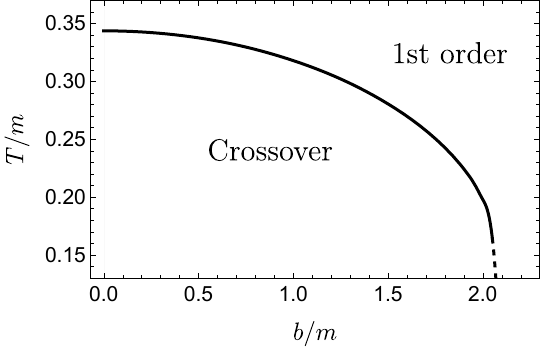}
    \caption{The phase diagram in the ($b/m,T/m$) plane. The solid black curves show the critical points splitting into the first-order phase transition and crossover regions. In the parameter region for small $T/m$ and large $b/m$, denoted by the dashed curve, the reconnection takes place before the second-order phase transition undergoes.}
    \label{fig:phase}
\end{figure}

Lastly, we show the phase diagram of our current-driven phase transitions in the ($b/m,T/m$) plane in figure \ref{fig:phase}.
The solid black curve shows the critical points ($b_c /m, T_c /m$) splitting into the region where the first-order phase transitions and crossover undergo.
In the limit of $b=0$, the critical temperature approaches $T_c /m \approx 0.3436$ as studied in \cite{Nakamura:2012ae,Matsumoto:2018ukk}.
It turns out that the axial gauge potential $b$ shifts the critical point to a lower temperature.
In the parameter region for small $T/m$ and large $b/m$, denoted by the dashed curve, the current-driven phase transitions are not well-defined since the reconnection we observed at zero temperature takes place.
Note that the phase diagram in the ($m/b,T/b$) plane can be obtained by replacing the parameters, but is not shown here because there is no qualitative difference compared to figure \ref{fig:phase}.

\section{Discussion on reconnection}\label{sec:disc}

In this section, we address the interpretation of a distinctive ``reconnection" phenomenon, observed in section \ref{sec:non-linearity}, on the analogy of band structure and discuss its possible implications for real WSMs.

Consider the energy bands in a WSM as two highways running parallel. In standard semimetals, these highways never intersect, maintaining a gap representing the energy difference between the valence and conduction bands. This gap is crucial for the semimetal's electronic properties, dictating electron mobility and conductivity.
In WSMs, however, a remarkable transformation occurs due to strong spin-orbit coupling and specific material symmetries: these parallel highways bend towards each other, eventually intersecting at points known as Weyl nodes. At these nodes, electrons acquire high mobility, akin to being massless, allowing for exceptional conductivity \cite{PhysRevX.5.031013,Guo:2022dpc,Wang2018,PhysRevB.96.174205}.

The phenomenon of reconnection in WSMs can be visualized as a significant infrastructural change on these highways. 
Initially separate, the introduction of interactions, such as electron-electron interactions or external fields, causes these highways to merge, forming new pathways.
Interestingly, our observation in Figure \ref{fig:JEplot} indicates that the forming of new pathways is gradually progressed due to the competing effect quantified by $m/b$, and eventually leads to a critical juncture where reconnection takes place, indicating a pivotal moment in the dynamic process.
This reconnection signifies a substantial alteration in the band structure, leading to a reconfiguration of Weyl nodes and, consequently, novel electronic properties.
This intuitive analogy simplifies the complex quantum mechanics underlying WSMs but effectively captures the reconnection behavior's essence and its potential impact on the material's properties. 

The insights gained from our holographic model provide a qualitative understanding of the behavior in real WSMs.
In actual WSMs, such reconnection could precipitate a dramatic shift in electrical conduction. 
The reconnection observed could translate to abrupt changes in transport properties in actual materials, such as a surge in conductivity or the emergence of novel quantum oscillations.
Experimental verification through magneto-transport measurements or alterations in the density of states near the Fermi level would be pivotal in confirming the phase transition.



\section{Conclusion and remarks}\label{sec:con}
In this study, we delved into the intricate non-linear current-electric field ($J$-$E$) response within the holographic Weyl semimetal (WSM) as modeled by the D3/D7 framework with $U(1)$ gauge fields. By introducing the axial gauge potential through the spiral configuration of the D7-brane in the bulk setup, we unraveled the non-Ohmic conductivity behavior that reflects the presence of strong interactions embedded in the WSM. Our findings suggest that the transport characteristics of the system are influenced by interactions operating beyond linear regimes.

At zero temperature, our analysis uncovered a remarkable ``reconnection" phenomenon inherent to the holographic WSM, indicative of quantum phase transition characterized by a significant alteration in the $J$-$E$ relationship. This transformative event hints at a reorganization of the Weyl nodes, thereby inducing shifts in the topological properties of the system. Further exploration of the non-linearity in the $J$-$E$ diagram at finite temperatures led us to discern critical exponents that exhibit correspondence with Landau theory principles governing equilibrium phenomena. This alignment not only emphasizes the importance of our research findings but also establishes a robust connection with existing theoretical frameworks

Before concluding the paper, we make some remarks.
Our study of the critical phenomena of the current-driven phase transitions reveals that the axial gauge potential does not change the critical phenomena studied in the ordinary metallic system \cite{Nakamura:2012ae} as well as it may be regarded as one of the thermodynamic variables.
While we strict ourselves to compute the critical exponents ($\beta,\delta$), the other critical exponents are expected to agree with the mean-field values as studied in \cite{Matsumoto:2018ukk} due to the qualitative similarity.

Another interesting direction is to investigate the dynamical stability of the steady state in our holographic WSMs.
The steady-state region exhibiting the negative differential conductivity is known to be dynamically unstable, which suggests that small perturbations could lead to significant deviations from the steady state \cite{Ishigaki:2021vyv}. This instability is critical for understanding the transport properties of the system and could have implications for the development of devices that exploit non-linear conductivity phenomena.
In our holographic WSMs, the other steady-state branch originated from the WSM phase is emerged as shown in figure \ref{fig:JEplot}.
It would be interesting to study the dynamical stability and emergence of inhomogeneous branches in our setup.

It would be intriguing to extend our study of holographic WSMs to non-relativistic backgrounds, such as Lifshitz or Schrödinger geometries. This novel approach would allow us to explore the response of WSMs under varied conditions, potentially unlocking new possibilities in advanced material science and quantum information technologies. The investigation of non-relativistic backgrounds could provide valuable insights into the behavior of WSMs in more realistic and diverse scenarios.

\section*{Acknowledgement}
M.~M.~is supported by Shanghai Post-doctoral Excellence Program (No.\,2023338). 

\bibliography{thebib}

\end{document}